\documentclass{aa} 
\usepackage{booktabs}
\usepackage{graphicx}
\usepackage{txfonts}
\usepackage{changes}
\usepackage{textcomp}
\usepackage{gensymb}

\def\comment#1{\relax}

\usepackage{color}

\begin{document} 

\title{{Traces of} wobbling accretion disk in X-ray pulsar Her X-1 from observations of the ART-XC telescope of the SRG observatory.}
    \titlerunning{Precessing and wobbling accretion disk of the Her X-1 from \textit{SRG}/ART-XC observation}
    \authorrunning{V.~M.~Revnivtsev et al.}

   \author{ V.~M.~Revnivtsev\inst{1,2}\thanks{vlad.revnivtsev@cosmos.ru},
            K.~A.~Postnov\inst{1},
            S.~V.~Molkov\inst{2}, 
            N.~I.~Shakura\inst{1},
            A.~A.~Lutovinov\inst{2},
            I.~Yu.~Lapshov\inst{2},
            D.~A.~Kolesnikov\inst{1},
            A.~Yu.~Tkachenko\inst{2}
        }

   \institute{$^1$ Moscow State University, Sternberg Astronomical Institute, 119234 Moscow, Russia\\
            $^2$ Space Research Institute of the Russian Academy of Sciences, 117997 Moscow, Russia\\
            }

    \date{\today}
 
\abstract{Long uninterrupted observations of the X-ray binary system Her X-1 were performed
with the Mikhail Pavlinsky ART-XC telescope of the Spectrum-R\"ontgen-Gamma (SRG) X-ray
Observatory in the 4--25 keV energy range with a total exposure of about two days around
the main turn-on of the X-ray source. We present the results of timing and spectral analysis
of these observations. The opening of the X-ray source is determined to occur at the orbital
phase $\phi_\mathrm{b}\approx 0.25$. The analysis of the X-ray light curve reveals a first
direct observational evidence of the nutation of a tilted  precessing accretion disk with
a period of $\simeq0.87$ days. The appearance of X-ray pulsations near the orbital
phase $\phi_\mathrm{b}\simeq 0.77$ prior to the main turn-on at the maximum of the
nutation variability has been also detected. During the X-ray eclipse, a non-zero X-ray
flux is measured, which is presumably associated with scattering of an X-ray emission
in a hot corona around the optical star illuminated by the X-rays from the central
neutron star. An increase in the X-ray flux after the main turn-on can be described
by the passage of the radiation from the central source through a scattering corona
above the precessing accretion disk.}

\maketitle
%

\section{Introduction}
Her X-1/HZ Her is an eclipsing close X-ray binary system harboring a neutron star with
a mass of $\approx 1.5{M_\odot}$ and the optical star HZ Her with
a mass of $\approx 2{M_\odot}$. The HZ Her was known as a variable star with an unusually
large optical variability. After the discovery of the X-ray pulsar Her X-1 and its optical
identification with HZ Her, it became clear that it enters a binary system consisting
of a normal star and a neutron star \citep{1972ApJ...174L.143T, 1973ApJ...184..227G}.   
The large optical variability of HZ Her is a consequence of the heating of the normal
star by the X-ray radiation of the neutron star 
\citep[the reflection effect,][]{1972IBVS..720....1C, 1976ApJ...209..562G, 2021A&A...648A..39S}.

An X-ray flux from Her X-1 demonstrates several characteristic periodicities: X-ray pulsations
with the neutron star's spin period of $P_\mathrm{x}\approx 1.24$ s; the variability with
the binary orbital motion with a period of $P_\mathrm{b}\approx 1.7$ days; the superorbital
variability with a period of $P_\mathrm{prec}\approx 35$ days. The 35-day period is caused
by the precession of an accretion disk around the neutron star in the retrograde direction
to the orbital motion. Because of the large orbital inclination of the binary
system $i\approx 90$ degrees \citep{1974ApJ...191..483C}, the central X-ray source is
periodically eclipsed both by the optical star and the inclined accretion disk.
The precession period consists of two X-ray bright states: a bright main-on state
(lasting about seven orbits) followed by a low state with almost zero X-ray flux
(lasting about four orbits), and a shorter secondary short-on state with lower X-ray
flux than in the main-on state (lasting about five orbits) followed by the second low
state (about four orbits). A detailed description of the phenomenology of the 35-day
X-ray variability can be found in
\citet{1998MNRAS.300..992S, 2020ApJ...902..146L,2021A&A...648A..39S}.
The X-ray luminosity of the source in the main-on state reaches $L_x\approx 10^{37}$ erg/s.

\begin{figure}[ht!]
    \centering
    \includegraphics[width=\columnwidth]{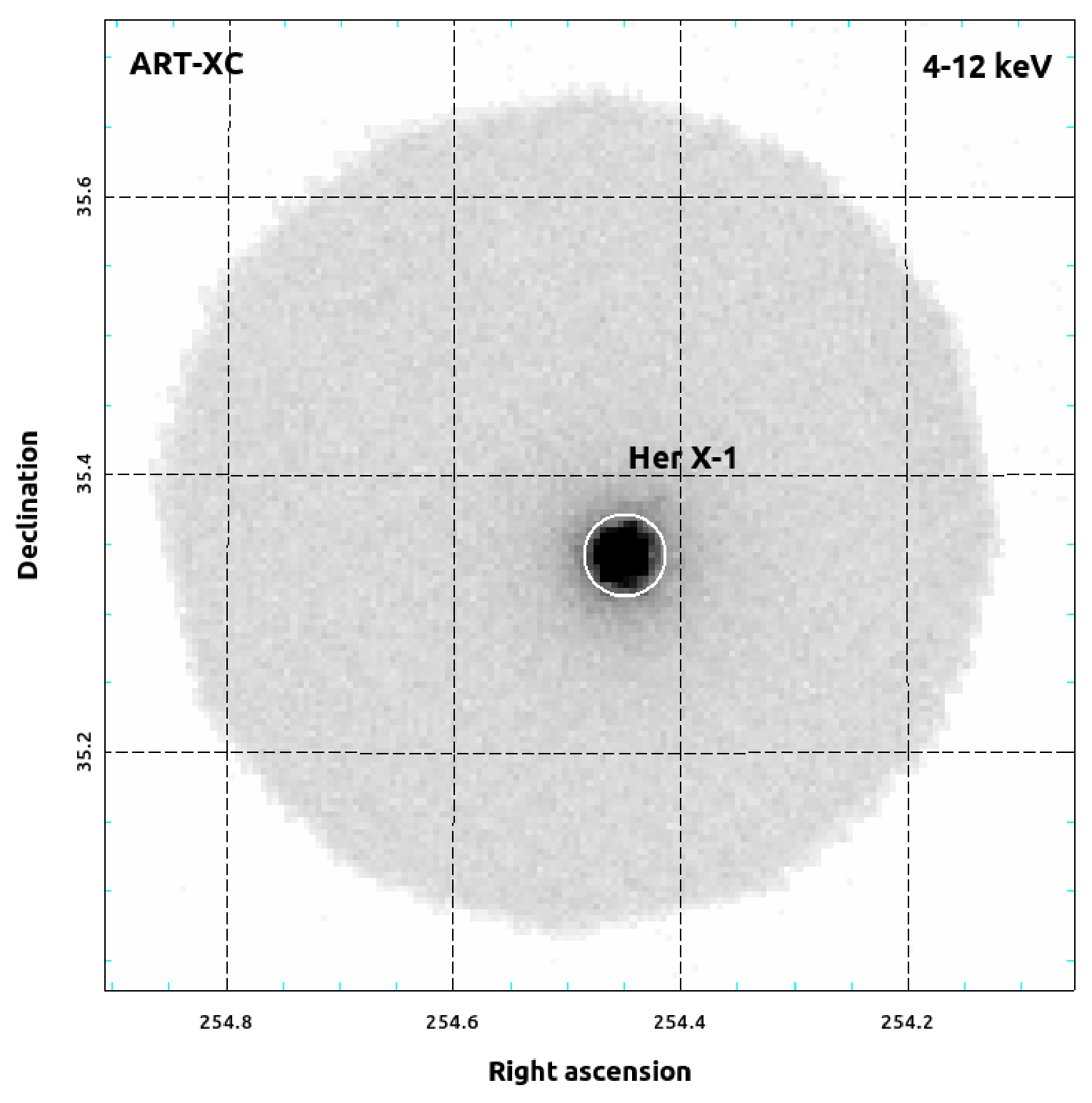}
    \caption{ART-XC 4--12 keV image of the sky region around Her X-1.}
\label{fig1}
\end{figure}

The accretion disk is the key element of the Her X-1 binary system that modulates
the observed X-ray radiation. The shape and dynamics of the disk can be caused by
the reaction of the asymmetrical coronal disk wind \citep{1994A&A...289..149S},
radiation reaction from the central luminous X-ray source \citep{1999MNRAS.308..207W}
or, alternatively, by a combination of tidal effects in the binary system and dynamical
effects of accreting gas stream leading to the precession of the disk with a period of
35 days \citep{1998MNRAS.300..992S}.

As shown in \citet{2000ApJ...539..392S}, phenomenologically, precession of a twisted
tilted accretion disk around non-precessing neutron star is capable of explaining X-ray
pulse profile changes in Her X-1.  
Alternatively, X-ray pulse profile changes can be explained by a model of freely precessing
neutron star \citep{2013MNRAS.435.1147P}. Recently, evidence for triaxial neutron star
precession was found \textit{Fermi}/GBM measurements of pulse period \citep{2022MNRAS.513.3359K}.
Based on the RXTE observations, a model of a wobbling (nutating) accretion disk was
proposed by \citet{2006AstL...32..804K} to explain an anomalous appearance of the X-ray
source during its low states, when the neutron star becomes visible for a short time
to the observer.

Her X-1 has been observed many times by different X-ray observatories in various states
\citep[see e.g.][and references therein]{1972ApJ...174L.143T, 2000AstL...26..691L, 2008A&A...482..907K, 2020ApJ...902..146L, 2022AAS...24024603C, 2024Univ...10..298L}.
An extended hot corona above the accretion disk was detected during X-ray eclipses of
Her X-1 in RXTE observations \citep{2015ApJ...800...32L}.
In the first two all-sky surveys by the SRG/eROSITA telescope (0.4--10 keV,
\citep{2021A&A...647A...1P}), the binary system was observed (about once every four hours
for $\sim 40$ s) during the low states, {where the neutron star is screened from the
observer by the accretion disk}, with a full exposure of 570 and 588 s accumulated over
several days \citep{2021A&A...648A..39S}. The SRG/eROSITA observations revealed an
orbital modulation of the soft X-ray flux caused by scattering on a hot corona above
the disk and on a hot optically thin corona above the X-ray-heated atmosphere of HZ Her. 

This paper presents the results of dedicated continuous observations of the X-ray source
Her X-1 by the Mikhail Pavlinsky ART-XC telescope of the SRG observatory in the 4--25 keV
energy range during $\approx 1.94$ days near the main X-ray source turn-on. The observations
captured the end of the second low state, the X-ray eclipse, the main turn-on, and
the X-ray flux increase in the main-on state beginning. These phases of the 35-day
period are used to search for the expected modulation of the X-ray flux during nutation
oscillations of the disk in the low state and for study features of the flux increase
after the source's turn-on. 

\section{ART-XC Telescope}

The Mikhail Pavlinsky ART-XC telescope is one of two telescopes installed aboard
the Spectrum-R\"ontgen-Gamma X-ray observatory. The SRG observatory was launched
in 2019 from the Baikonur cosmodrome to the L2 libration point of the Sun-Earth
system \citep{2021A&A...656A.132S}. 

The ART-XC telescope consists of seven co-directional identical X-ray modules.
The mirror system provides a field of view of $\sim$ 0.3 square degrees. The operating
energy range of the ART-XC telescope is 4--30 keV, but the detectors can operate
up to $\sim120$ keV and the highest sensitivity is achieved at 4--12 keV. The time
resolution of the telescope is $\sim$ 23 $\mu$s and the angular resolution
is $\sim 53$\text{"}. Such parameters of the telescope make it possible not only to
survey large areas of the sky, but also to carry out a detailed spectral and temporal
analysis of the radiation from individual objects \citep[see][for a more detailed
description of the instrument]{2021A&A...650A..42P}.

\section{Observations and data analysis}

Dedicated observations of Her X-1 were carried out by the SRG/ART-XC telescope on
August 25--26, 2023, near the moment of the main turn-on during 541th precession cycle
according to the ephemerid \citep{2013MNRAS.435.1147P}
\begin{equation}
T_\mathrm{MJD} = 42409.84 + 20.5 \times P_\mathrm{orb} \times (N_{35} - 31), 
\end{equation}
where $N_{35}$ is the precession cycle number, $P_\mathrm{orb}$ is the orbital period. 

The ART-XC telescope data were processed using standard software {\sc artproducts v0.9}
with the calibration data version {\sc CALDB} 20230228; the photon registration times
were adjusted to the solar system barycenter.

Figure \ref{fig1} shows an image of the sky region around Her X-1 obtained with
ART-XC telescope over the entire observational period. For further analysis only photons
fell inside a circle with radius of 1\farcm8 around the source were used. The background
was estimated by collecting photons from a circular region with a radius of 3\farcm6 on
the detectors that did not overlap with the region in which the source photons were
extracted.

\begin{figure}[ht!]
\centering
\includegraphics[width=\columnwidth]{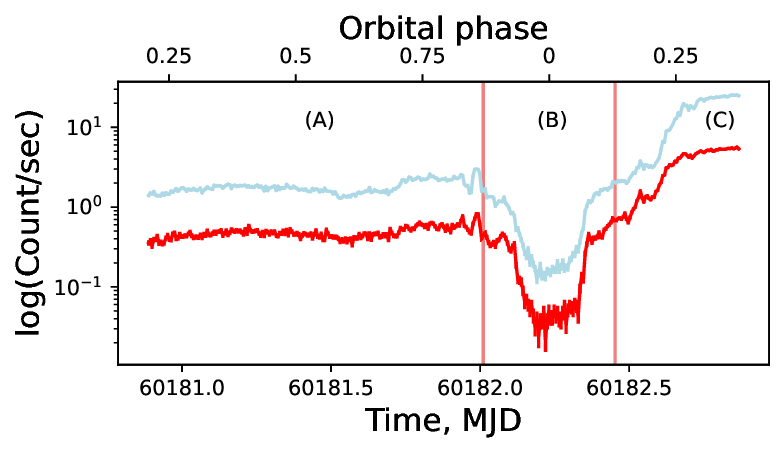}
\caption{ART-XC X-ray light curves of Her X-1 in two energy bands with a time resolution
of $500$ s (4--12 keV in blue and 12--25 keV in red). Zero orbital phase corresponds to
the neutron star eclipse by the optical component.}
\label{fig2}
\end{figure}

Figure \ref{fig2} shows the light curves of Her X-1 in two energy bands (4--12 keV, 12--25 keV)
with a time resolution of 500 s. The figure shows that the entire X-ray light curve can be
roughly divided into three main parts: (A) the pre-eclipse part, when the system is in
the off (low) state and the neutron star is screened from the observer by the accretion
disk; (B) the eclipse of the X-ray source by the optical star, where the observed X-ray
flux reaches its minimum; (C) the post-eclipse part, when the source is opened to the
observer. These stages will be discussed in more details below.

\begin{figure*}[ht!]
\centering
\includegraphics[width=\columnwidth]{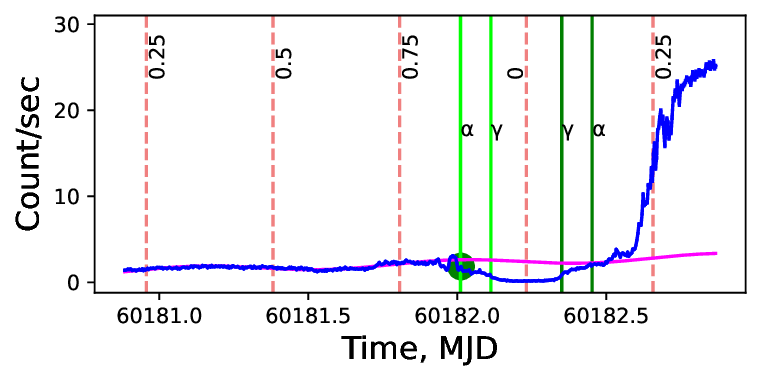}
\includegraphics[width=\columnwidth]{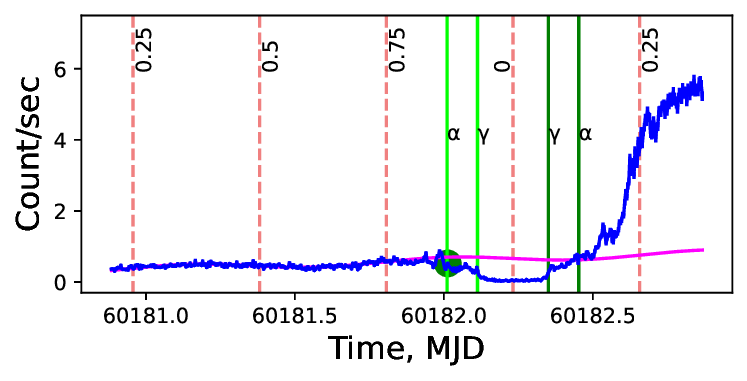}
\includegraphics[width=\textwidth]{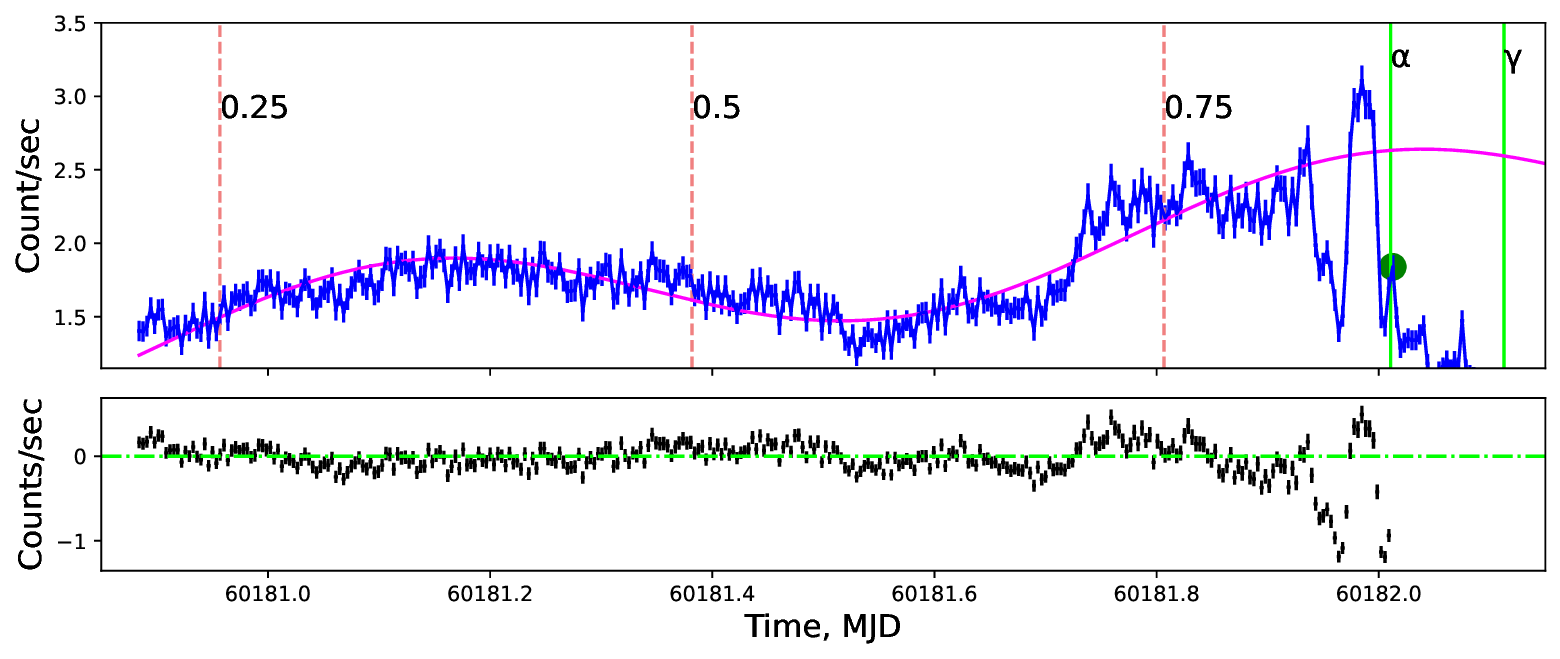}
\includegraphics[width=\textwidth]{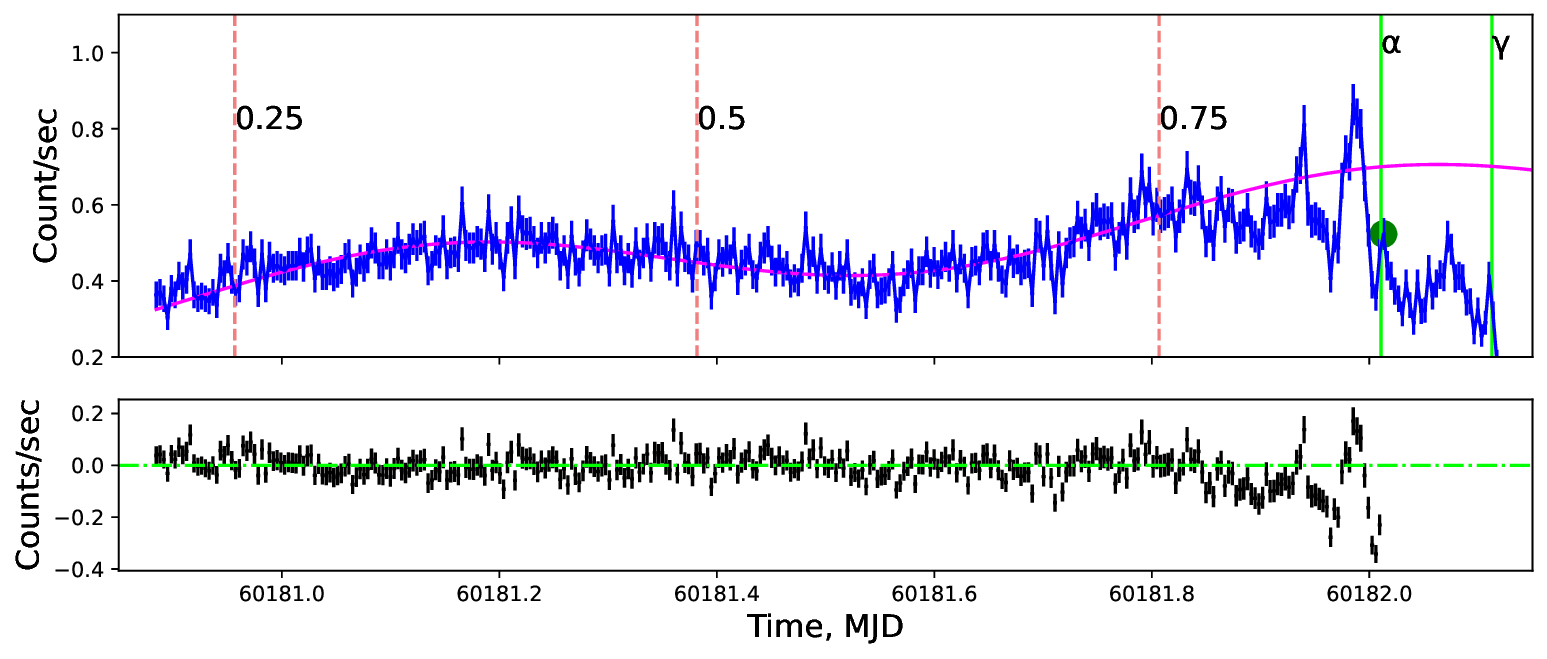}
\caption{ART-XC 4--12 keV (top left) and 12--25 keV (top right) light curves; enlarged
pre-eclipse part of the 4--12 keV light curve (center) and enlarged pre-eclipse part
of the 12--25 keV light curve (bottom). Blue errorbars correspond to $300$s data binning,
pink line corresponds to the best fit of the model parameters (3), and green solid circles
correspond to the orbital phase up to which the model describes the observed light
curve (the beginning of X-ray eclipse ingress). Light green vertical lines mark
specific eclipse phases: $\alpha$ -- the beginning of the accretion disk eclipse,
$\gamma$ -- the neutron star eclipse. Dark green vertical lines mark the end of the
corresponding eclipse phases. Red dashed vertical lines show specific orbital phases.
Black errorbars show the model residuals.}
\label{fig3}
\end{figure*}

\begin{table}[ht]
  \caption{Best-fit parameter of the nutation model (3).}
  \centering%
  \resizebox{\columnwidth}{!}{
    \renewcommand{\arraystretch}{1.5}
    \begin{tabular}{|c|c|c|c|c|c|}\hline
      Energy range, keV& $a_0$, counts/s  & $\phi$ &  k, counts/s/d & b, counts/s\\\hline 
      4--12 & $0.38\pm0.01$ & $1.49\pm0.02$ & $0.85\pm0.01$ & $1.297\pm0.008$\\\hline
      12--25 & $0.088\pm0.003$ & $1.6\pm0.04$ & $0.24\pm0.01$ & $0.344\pm0.004$\\\hline
    \end{tabular}
  }
  \label{tab:tbl1}
\end{table}

As seen from Figure \ref{fig2}, the light curves are similar in both energy ranges,
so they can be exemplified by  the 4--12 keV light curve (where ART-XC has a larger
effective area than in the 12--25 keV range). The features on the 12--25 keV light
curve generally coincide with those in the 4--12 keV range, except cases that will
be considered separately. The orbital phases on the top of Figure \ref{fig2} are
calculated using ephemerids from \citep{2009A&A...500..883S}.

Consider a geometrical model where the accretion disk centered on the neutron star
to be a  circle with radius $R_d \approx 0.7r_{L1}$ \citep{1977ApJ...216..822P}, 
where $r_{L1}\approx a(0.5+0.227\log_{10}q)$ is the distance between the compact
star and the inner Lagrangian point $L_1$ \citep{2002apa..book.....F},
$q=M_x/M_o\approx 0.7$ is the binary mass ratio in Her X-1
\citep{2008ARep...52..379A, 2014ApJ...793...79L}, $a$ is the binary orbital separation,
and the optical star is filling its Roche lobe $R_L(M_o)=af(\frac{1}{q})$,
where $f(q) =\frac{0.49q^{2/3}}{0.6q^{2/3}+\ln(1+q^{1/3})}$ \citep{1983ApJ...268..368E}.
Her X-1/HZ Her is an almost edge-on binary system, so in this geometrical model we can
estimate specific orbital phases of the beginning and end of eclipse of the accretion
disk  $\alpha$ and the neutron star $\gamma$ by the optical star:

\begin{equation}
\left\{
\begin{aligned}
    \gamma &= \pm\frac{1}{2\pi} \arcsin\frac{R_o}{a}\approx \pm 0.07 \\
    \alpha &= \pm\frac{1}{2\pi} \arcsin\frac{R_o + R_d}{a}\approx \pm 0.13\,.
\end{aligned}
\right.
\end{equation}
{(Note that these phases are almost insensitive to the assumed binary mass ratio $0.5<q<1$.)
In this model, the  beginning and end of eclipse of the accretion disk and neutron star,
are symmetric with respect to zero orbital phase (full eclipse of the neutron star).
Thus, the orbital phases 
$\alpha$, $\gamma$, are  0.87, 0.93, and $\gamma$, $\alpha$ = 0.07, 0.13, respectively
(see Figure \ref{fig3}, top panels). 
}

\section{Results}

\subsection{Spectral analysis}

Her X-1 spectra were constructed from ART-XC data using the {\sc XSPEC} model
{\sc highecut(powerlaw)+gaussian} \citep{2020A&A...642A.196S, 2013arXiv1309.5361F}.
Along the light curve the spectral parameters are found to remain approximately
constant: $E_{\rm cut}\sim18$ keV, $E_{\rm fold}\sim8$ keV, photon index $\Gamma \sim0.6$.
Almost constant spectral parameters in the range of 0.5-7 keV across the 35-day cycle were
also reported from the recent analysis of Her X-1 observations by AstroSat Soft X-ray
Telescope  \citep{2024Univ...10..298L}. However, a simple power-law model is sufficient
to describe the spectrum in the X-ray eclipse. Moreover, the spectrum immediately after
the eclipse up to the main turn-on of the source the spectrum becomes harder and an
additional soft component appears (see below).


\begin{figure}[ht!]
    \centering
    \includegraphics[width=\columnwidth]{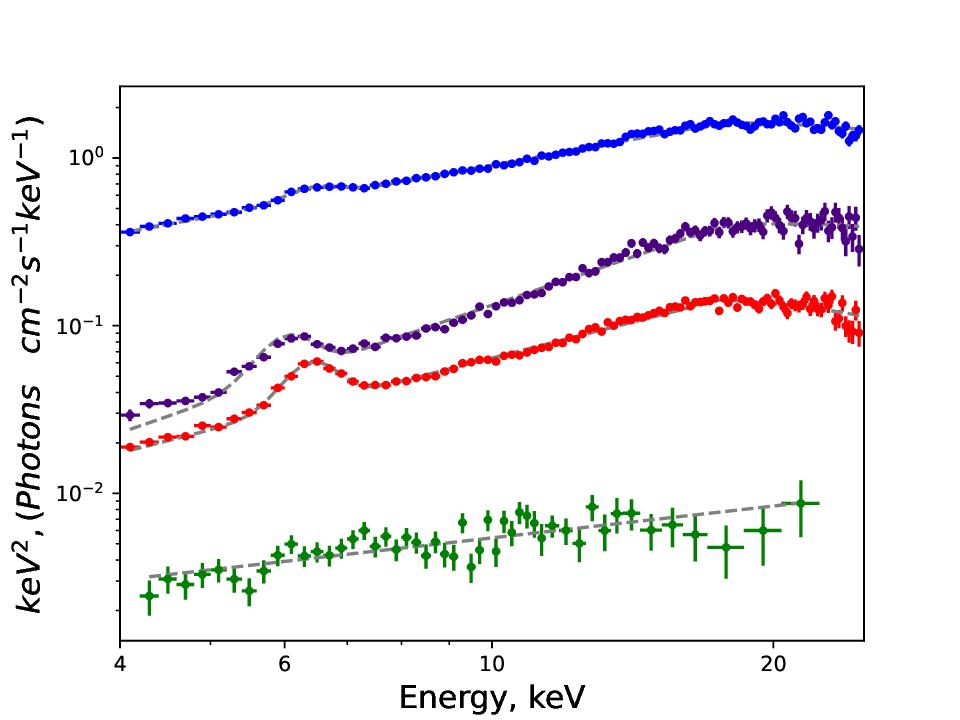}
    \caption{ART-XC Her X-1 spectrum approximated by the {\sc highecut(powerlaw)+gaussian}
model. From top to bottom: the main-on high state (MJD 60182.7 -- 60182.9, in blue);
the low state before the main turn-on  after the X-ray eclipse when the neutron star is
screened by the accretion disk (MJD 60182. 5 -- 60182.6, in purple);  the low state
before the X-ray eclipse (MJD 60180.9 -- 60181.5, in red);
the X-ray eclipse (MJD 60182.15 -- 60182.3, in green).}
\label{fig4}
\end{figure}

Figure \ref{fig4} shows X-ray spectra in the X-ray eclipse (MJD 60182.15 -- 60182.3, in green),
in the pre-eclipse phase (MJD 60180.9 -- 60181.5, in red), and in the main-on high state
(MJD 60182.7 -- 60182.9, in blue). In the post-eclipse phase before the main turn-on
the spectrum of a short hard peak (MJD 60182.5 -- 60182.6) an additional soft component
is required (in purple). 

\subsection{Analysis of post-eclipse X-ray spectrum }

\begin{figure}[ht!]
  \centering
  \includegraphics[width=\columnwidth]{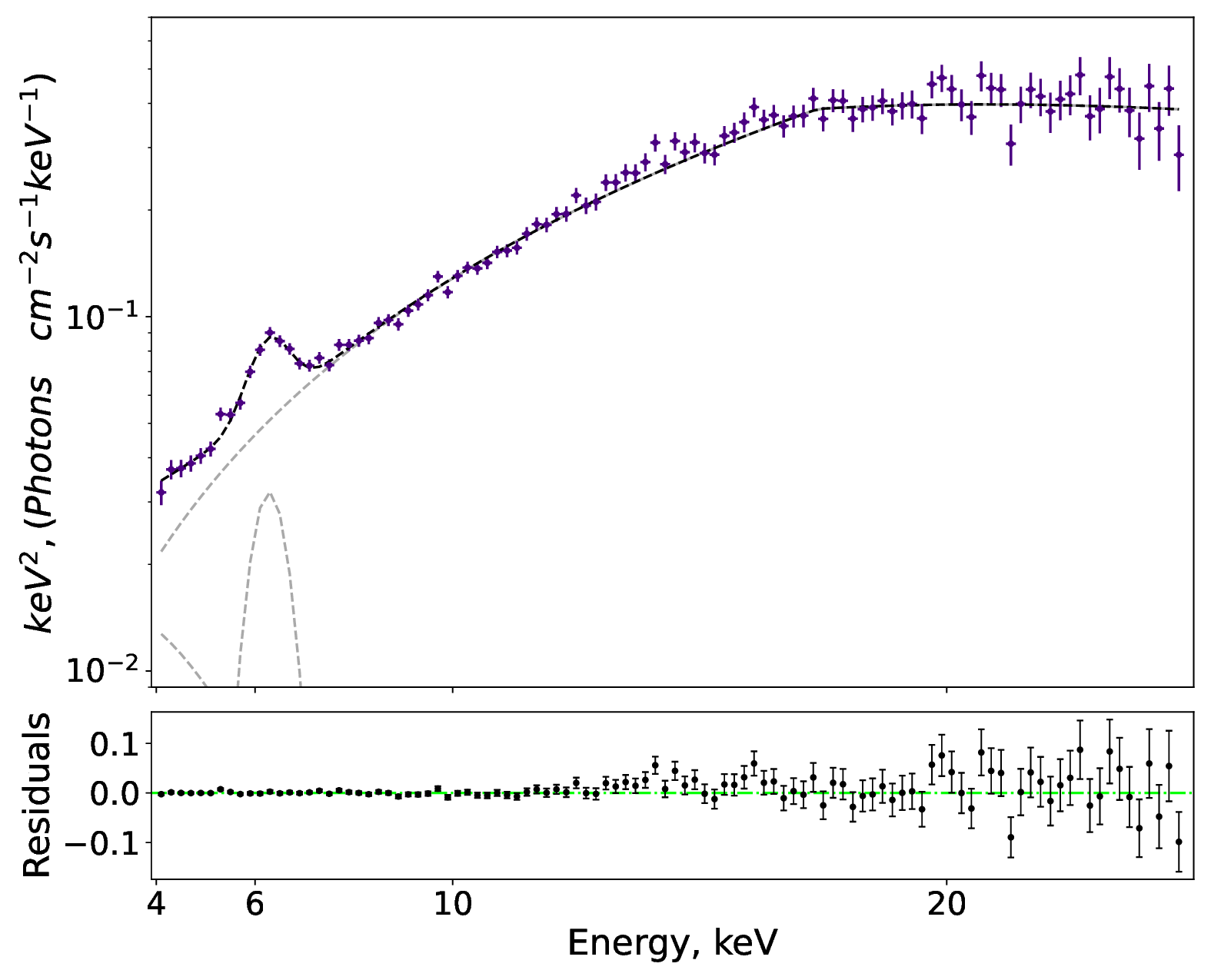}
  \caption{(Top) 
    Purple points -- a hard spectrum in the low state (see Figure \ref{fig4}). Gray
    dashed lines indicate models: {\sc highecut(powerlaw), gaussian, blackbody}.
    The black dashed line is the total model. (Bottom) The difference between
    the observation points and the total model.}
  \label{fig5}
\end{figure}

As mentioned above, for a short time (MJD MJD 60182.5 -- 60182.6)
after the X-ray eclipse before the main turn-on, an additional soft component,
not described by the {\sc highecut(powerlaw)+gaussian} model  appears in the spectrum.
This soft component can be described by the model of a black body with
the temperature of $kT \sim 0.8$ keV (see Figure \ref{fig5}). At a source distance
of $\sim6.6-6.1$ kpc {\citep{1997MNRAS.288...43R,2014ApJ...793...79L}} the observed
soft  excess corresponds to radiation from a spot with a size of $\sim 1$ km

\begin{table}[ht]
  \caption{{Model parameters of the ART-XC spectrum of Her X-1 after eclipse before
      the main-on (in purple in Fig. 4 and 5). }}
  \centering
  \begin{tabular}{@{}lll@{}}
    \toprule
    Parameter & Value  & Error \\ \midrule
    $E_{\rm cut}$ (keV) & 17.41 & 0.45 \\
    $E_{\rm fold}$ (keV) & 10.44 & 1.09 \\
    $\Gamma$ & 0.01 & 0.05 \\
    norm$_{\rm pow}$ & 0.0013 & 0.0002 \\
    $E_{\rm Fe}$ (keV) & 6.23 & 0.07 \\
    $\sigma_{\rm Fe}$ & 0.4 & fixed \\
    norm$_{\rm Fe}$ & 0.0008 & 0.0001 \\
    kT & 0.79 & 0.22 \\
    norm$_{\rm bb}$ & 0.0004 & 0.0002 \\ \bottomrule
  \end{tabular}
  \label{tab:tbl2}
\end{table}

The emergence of this soft component for a short time before the main turn-on of Her X-1
can be related to the contribution of the hot spot from the interaction region between
the gas stream from the inner Lagrangian point and the outer parts of the disk.
The interaction of the infalling matter with the outer disk
at $R_{out}\sim1.6 \times 10^{11}$cm produces shock with a  temperature behind
the front $kT \sim \frac{3}{16}m_pV^2$ ($m_p$ is the mass of a proton). At characteristic
ballistic velocities of the  stream $V\sim600\frac{km}{s}$, this temperature is close to
the temperature obtained from the spectral data.

\subsection{Analysis of the pre-eclipse light curve}

In this section, we consider the pre-eclipse part of the light curve in the low state,
when the emission from the central source is screened by the accretion disk.
The residual X-ray flux during the eclipse can be due to scattering of the X-ray emission
from the neutron star on the corona above the X-ray illuminated atmosphere of the optical
star and on the disk corona as suggested by the analysis of the SRG/eROSITA observations
of Her X-1  \citep{2021A&A...648A..39S}. It is seen in Figure \ref{fig2} that there is
a marked increase in the flux near the orbital phase of 0.7 before the eclipse followed
by the eclipse ingress. 

The observed  smooth  variability of the X-ray flux in the off-state before the main-on
seen in Fig. 3 may be tentatively related to the change of the disk tilt (wobbling)
due to tidal precession  \citep{2006AstL...32..804K}. While seen only once, the
characteristic time and phase of the observed variability are very close to the expected
values for the disk wobbling model. 
The nutation period of the disk is 
$\frac{1}{P_\mathrm{nut}} = 2 \times \frac{1}{P_\mathrm{b}} - \frac{1}{P_\mathrm{pr}}$.
For Her X-1 $P_\mathrm{nut}=0.87$ days, or  
in units of cyclic frequency $\omega_\mathrm{nut} \approx 7.2$ rad/d.
In the wobbling disk model, the pre-eclipse part of the light curve should be modulated
with the nutation period, with an additional term responsible for a much longer disk
precession, which can be considered linear in a timescale of several days:
\begin{equation}
F = {a_0}\times\sin( t\times\omega_\mathrm{nut} - \phi) + k\times(t-T_0) + b\,.
\end{equation}

Here $t$ is the time at a given moment, $T_0$ is the time of the beginning of observations
(fixed parameter), $\omega_\mathrm{nut}$ is the nutation frequency, $\phi$ is the phase of
nutation oscillations, $b$ is the parameter specifying the flux in the zero phase of
the modulation, $k$ is the linear slope coefficient of the precession component, $a_0$
is the amplitude of the nutation modulation.

In the middle panels of Figure \ref{fig3} we show the best approximation of
the pre-eclipse part of the Her X-1 X-ray light curve with parameters given in
Table \ref{tab:tbl1}. As can be seen in the bottom panels, there is a significant
decrease in the registered emission before the eclipse, starting at about orbital phase 0.75,
compared to the theoretical sine-like curve (purple line). Such a decrease in the flux
before the eclipse ingress can be explained by the gas stream crossing the line of sight
(the so-called pre-eclipse dip). As shown in \citet{1999A&A...348..917S}, the pre-eclipse
dip begins around orbital phase 0.8 in the precession phase before the main turn-on and
smoothly transits into the X-ray eclipse ingress (see also \citet{2011ApJ...736...74L}
for analysis of dips from the entire RXTE/PCA observations).

We also note that starting at about orbital phase 0.65 the 4--12 keV flux behaves less smoothly
than the 12--25 keV flux. Near orbital phase 0.75 small quasi-periodic fluctuations of
the 4--12 keV flux are observed with the characteristic time of $\approx 2000-3000$ s,
which roughly corresponds to the Keplerian time of the orbit with a radius of $10^{10}$ cm
around the neutron star. Since such pronounced peaks are not observed in the hard band,
we can assume that {such fluctuations of the X-ray emission are associated with the emission
of a bright spot in the region of the accretion stream  collision with the inner regions
of a tilted disk \citep{1999A&A...348..917S}}. 


Our analysis suggests that the sine-like modulation observed by ART-XC in
the pre-eclipse part of the Her X-1 X-ray light curve in the low state may be well explained
by the wobbling disk model. {Should this modulation be found persistent from longer
observations (for example, phase-connected in both off-states), they would provide
a direct observational manifestation of the nutation of the precessing accretion disk
in the Her X-1/HZ Her system}. 

\subsection{Analysis of the X-ray eclipse}

\begin{figure}[ht!]
  \centering
  \includegraphics[width=\columnwidth]{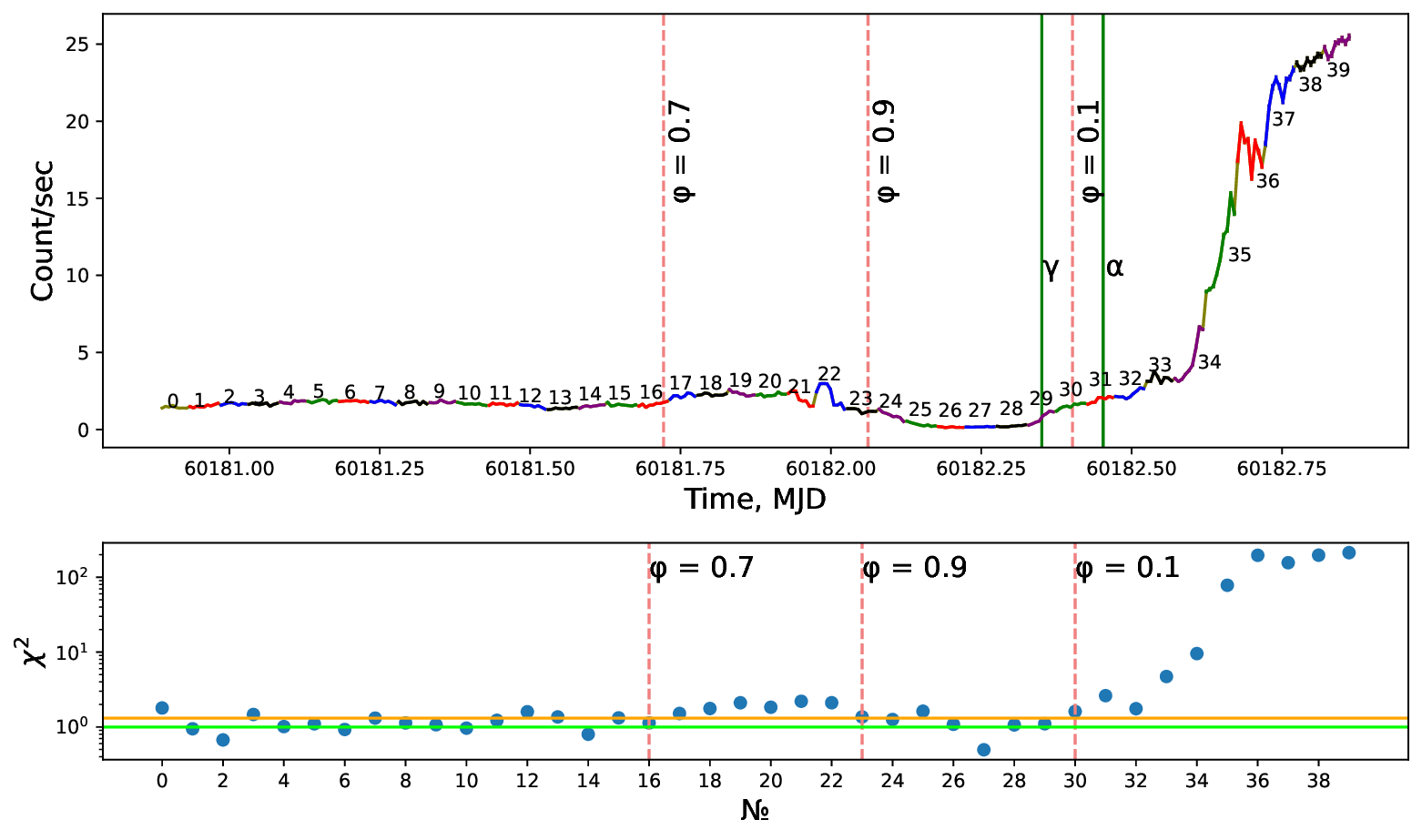}
  \caption{Upper panel: The ART-XC 4--12 keV light curve of Her X-1  divided into 40 equal time intervals. Green lines mark the specific eclipse phases: $\gamma$ = 0.07 is the  opening of the neutron star by the optical star, and $\alpha$ = 0.13 is the opening of the entire accretion disk by the optical star (see text). Bottom panel: Dependence of the $\chi^2$ value of the deviations of the count rate in the X-ray pulse from the constant in each time interval; green line corresponds to $\chi^2$=1 and and orange line to $2\sigma$-deviation of the pulsed flux  from the constant.}
  \label{fig6}
\end{figure}

Analysis of RXTE X-ray eclipses in Her X-1 was made previously by \citep{2015ApJ...800...32L}.
In this section, the eclipsing part of the ART-XC light curve is considered.
Figure \ref{fig2} shows that  the X-ray emission at the zero orbital phase does not disappear
completely. In the analysis by \citep{2015ApJ...800...32L}, the residual emission in
the X-ray eclipse was explained by scattering on a corona above the accretion disk.
Here we consider an alternative explanation that this residual emission may be caused by
the presence of an X-ray halo around the observer-facing part of the optical star eclipsing
the X-ray source. Note that the light curve in the eclipse is asymmetric with respect to
zero orbital phase, suggesting a  complex shape of the tilted accretion disk producing
asymmetric X-ray illumination of the optical star. As a result of the non-uniform heating
of the the optical star atmosphere by the central X-ray source, the parts of the optically
thin scattering corona above the heated up atmosphere of the star, visible during
the X-ray eclipse ingress and egress, differ from each other.

To estimate the flux scattered by the optical star corona, we  assume a ring-like X-ray
halo around the optical star. Then we can estimate the ratio of the area of the ring-like
halo visible at the orbital phase 0 to that of the hot corona above the atmosphere of the
optical star visible at the orbital phase 0.5, which was found in \citep{2021A&A...648A..39S}.
Considering the ratio of the corona thickness to the optical star radius
$\frac{H}{R_\mathrm{opt}}\approx0. 15$, the ratio of the radiation scattered by the corona
to the total emission from an X-ray source {in the energy range up to 8 keV} near the orbital
phase 0.5 can be {estimated as} $\frac{F_\mathrm{scat}}{F_\mathrm{x}}\sim 0.03$
\citep{2021A&A...648A..39S}. The flux from the hot corona above the optical star at different
orbital phases depends on its apparent area, from which we obtain an estimate for the expected
flux ratio near orbital phase zero: $\frac{F_\mathrm{scat}}{F_\mathrm{x}}\sim7\times10^{-3}$. 

From the ART-XC data the observed ratio of the 4--12 keV flux during the X-ray eclipse near
the zero orbital phase to the high-state flux  is approximately
$\frac{F_\mathrm{scat}}{F_\mathrm{x}}\approx{(6.53\pm0.18)}\times10^{-3}$, which is in good
agreement with the above model estimates. 

\subsection{The main turn-on time of the source}

\begin{figure}[ht!]
  \centering
  \includegraphics[width=\columnwidth]{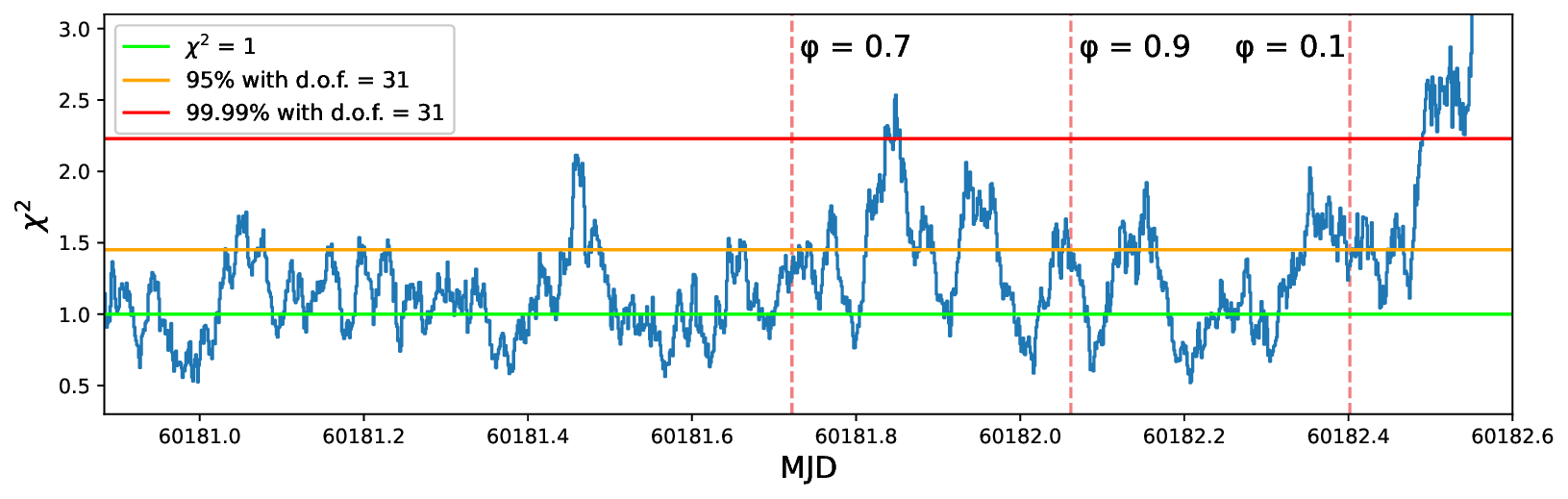}
  \caption{$\chi^2$ of the 4--12 keV pulse profile count rate deviations from constant constructed by the moving average method for time intervals of 4300 s with a 100 s shift. Green line corresponds to $\chi^2$=1, orange and red lines correspond to $2\sigma$ and $4\sigma$ flux deviations from constant, respectively.
  }
  \label{fig7}
\end{figure}

Figure \ref{fig6} shows that the main source turn-on -- a sharp increase in the flux from
Her X-1  -- occurs much later than the neutron star's exit from the eclipse by the optical
star (phase $\alpha=0.13$), which is expected in the geometrical model used. This means that
the neutron star is still screened from the observer by the tilted accretion disk after
the eclipse, and the turn-on happens near the orbital phase $\sim 0.2$.

The moment of the pulsations appearance when the source is opened by the precessing disk
is determined by the presence of a significant deviation of the pulse profile from constant.
To determine the moment of the turn-on, the 4--12 keV light curve was divided into short
equal-time intervals ($\approx 4300$ c), and in each time interval the count rates were
convolved with the pulsation period of the neutron star (Figure \ref{fig6}). The pulsation
period was determined from the last time interval (number 39 in Figure \ref{fig6}), when
the neutron star is fully opened to the observer and the pulsating fraction of the emission
is maximal, corrected for orbital motion:
$P_{\mathrm{spin}} = 1.2376986(1)$ s.

In the bottom panel of Figure \ref{fig6}, there are points exceeding $2\sigma$ deviation from
constant pulse profile inside the interval {(orange line)}. To test the significance of these
outliers, a similar plot was constructed using the moving average method (Figure \ref{fig7}).

\begin{figure*}[ht!]
  \centering
  \includegraphics[width=\columnwidth]{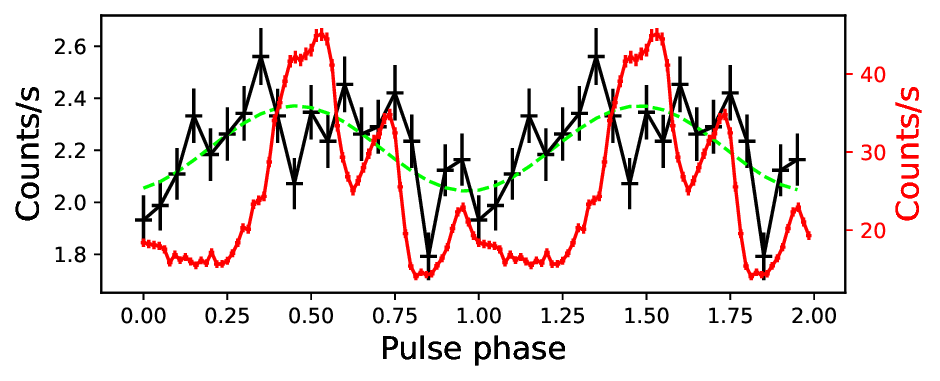}
  \includegraphics[width=\columnwidth]{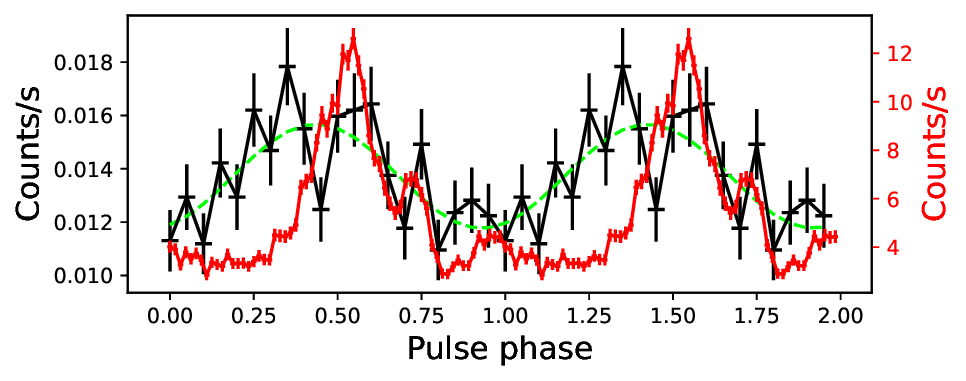}
  \caption{Pulse profiles (4--12 keV on the left, 12--25 keV on the right) near the orbital phase 0.77 (MJD = 60181.848 -- 60181.898; black color, scale on the left). Modeling of the pulse profile near the orbital phase 0.77 by sinusoidal curves (green color). For comparison, the pulse profiles after the main turn-on in time interval \#39 in Figure \ref{fig6} are shown (red color, scale on the right).
  }
  \label{fig8}
\end{figure*}

Figure \ref{fig7} clearly shows that significant pulsations beyond the $4\sigma$ (red line)
boundary was observed near the orbital phase $\approx 0.77$ before the eclipse by the optical
star. 
The observed  4--12 and 12--25 keV pulse profiles in the time interval
MJD = 60181.848 -- 60181.898 are shown in Figure \ref{fig8} (black dots). 

\begin{figure*}[ht!]
  \centering
  \includegraphics[width=\columnwidth]{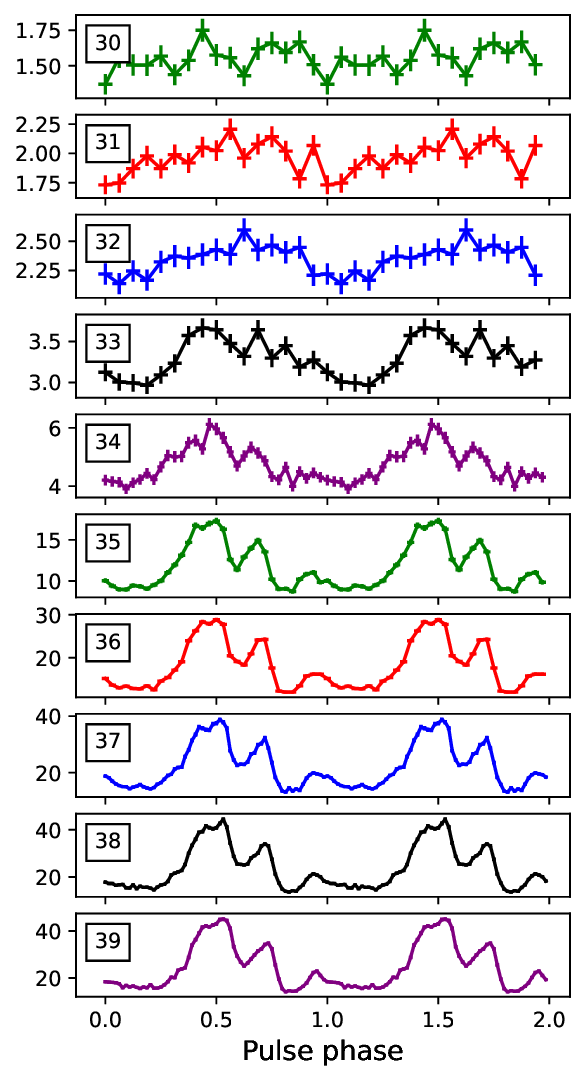}
  \includegraphics[width=\columnwidth]{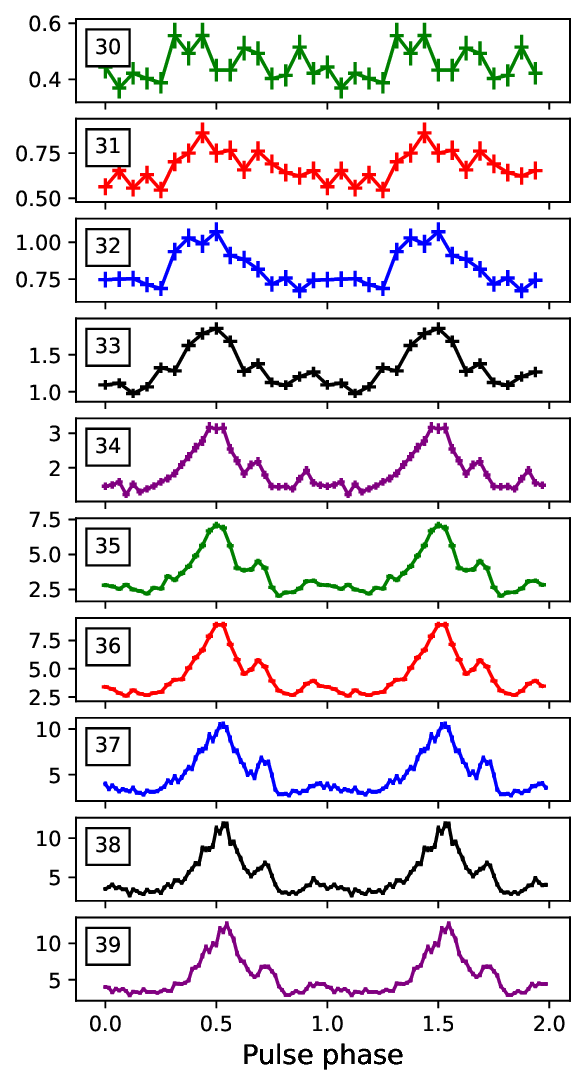}
  \caption{(Left) Pulse profiles in the post-eclipse 4--12 keV (left) and 12--25 (right) light curves. Time intervals are numbered as in Figure \ref{fig6}. 
  }
  \label{fig9}
\end{figure*}

\begin{figure*}[ht!]
  \centering
  \includegraphics[width=\columnwidth]{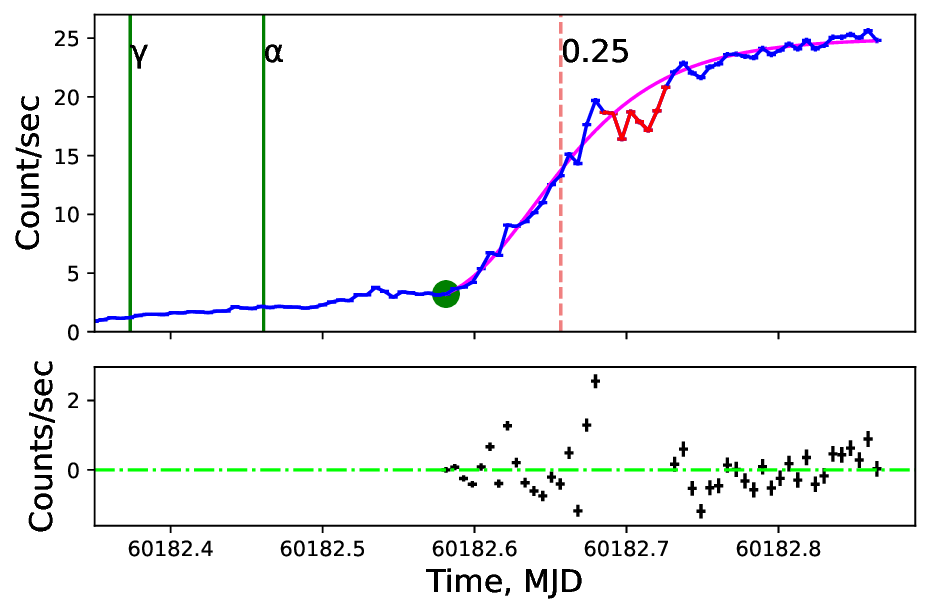}
  \includegraphics[width=\columnwidth]{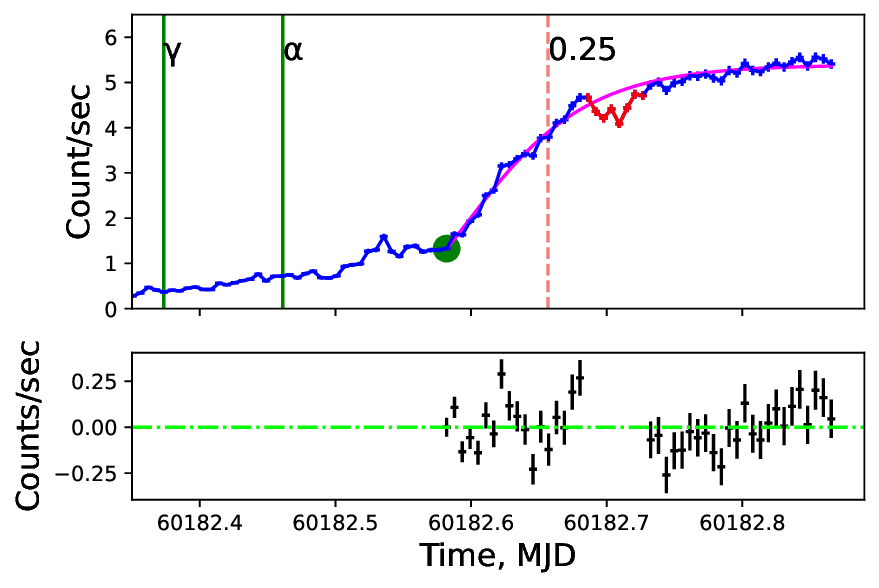}
  \caption{Modeling of the 4--12 keV (left) and 12--25 keV (right) flux on the exponential rise after the main Her X-1 turn-on, errorbars correspond to $500$ s data binning. Green filled circle marks the initial rise time of the X-ray flux $T_0$ (falls inside interval \#34 in Figure \ref{fig6}), red dots are ignored as they correspond to the X-ray flux dip caused by the expected inhomogeneous gas stream  from the Lagrangian point $L_1$. In the bottom panel, black dots with errorbars show difference between the observed and model count rate.
  }
  \label{fig10}
\end{figure*}

\begin{figure}[ht!]
  \centering
  \includegraphics[width=\columnwidth]{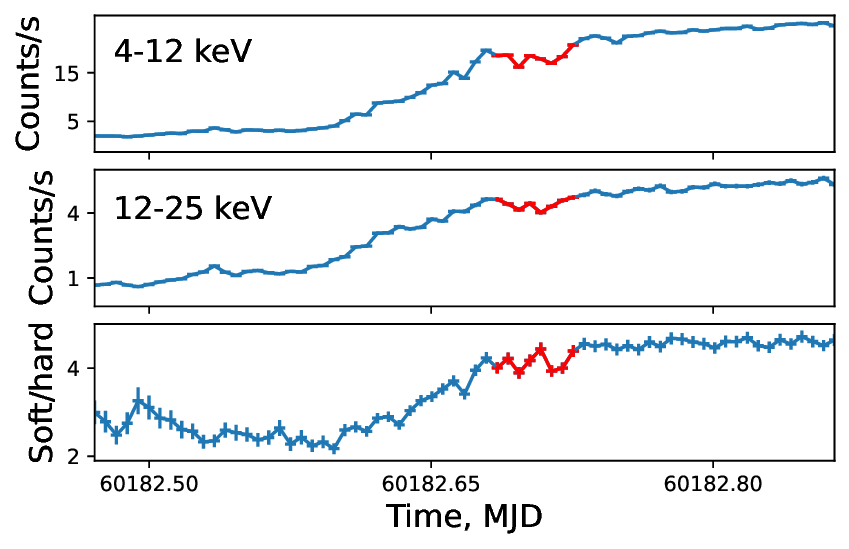}
  \caption{Top panel: the 4--12 keV light curve of the exponential flux growth. Middle panel: the corresponding 12--25 keV light curve. Bottom panel: softness of the X-ray light curve during the emission growth after the main turn-on; red dots mark the X-ray dip as in Figs. \ref{fig9}--\ref{fig10}. }
  \label{fig12}
\end{figure}

These profiles can be approximated by a characteristic sinusoidal trend
(green curves in Figure \ref{fig8}). The profiles itself seems to have a small shift relative
to the red pulse profiles obtained in the high state (interval \#39). This may be caused
by the fact that before the eclipse the observed pulse profile was not seen directly from
the neutron star, but, for example, represents reflected radiation on the tilted twisted
accretion disk, which leads to a quasi-sinusoidal shape of the profile with an additional
phase shift. 
However, we can not numerically estimate this phase shift or the distance to the reflecting
region of the twisted disk with sufficient reliability because of the very different shape
of the pulse profiles.

Thus we can conclude that the appearance of X-ray pulsations near the orbital phase 0.77 is
not the direct emission of the source, since the pulse profile is strongly blurred and strongly
affected by scattering on the corona of the accretion disk. Nevertheless, the fact that
the pulse signal was detected before the main turn-on is very remarkable because it falls
at the maximum of the sinusoidal flux fluctuation due to the disk nutation, which is
geometrically favorable for seeing the scattered emission from a neutron star still
covered by the disk at this precessional phase.

Figure \ref{fig9} shows the pulse profiles near the main turn-on in the post-eclipse part.
This figure clearly shows how the profiles change as the X-ray components are opened by
the optical star: in the interval \#30, where the neutron star is already open and the accretion
disk is partially open, no pulsations are seen; in intervals \#31--33, where the accretion
disk emerges from the eclipse, smoothed profiles similar to those observed before the eclipse
gradually appear; by intervals \#34--35, characteristic Her X-1 profiles appear. This interval
includes the moment of orbital phase 0.2, which we will consider to be the phase of the main
X-ray source turn-on. 

\subsection{Modeling of the exponential rise}

In this section, we discuss in more detail the time evolution of the X-ray flux in the post-eclipse
light curve. Following \citep{2005A&A...443..753K}, we consider that the X-ray radiation
from the neutron star reaches the observer, partially or completely absorbed and scattered
in the disk corona, differently in different energy ranges.

The exponential flux increase after the turn-on can be modeled assuming that the influence of
absorption is negligible, and the main role in the observed flux change is played by scattering
in the disk corona associated with the change in the number of particles along the line of sight
during the precessional motion of the disk{: $F(t) = {\tilde{F}} e^{-\tau(t)}$}. Taking into
account the observed very sharp dependence of the logarithm of the flux on time, to approximate
the X-ray light curve after the main turn-on we approximate the effective optical thickness by
a function exponentially depending on time $\tau(t) = A {e^{-\frac{t-T_0}{b}}}$ .
Then the flux rising is described as
\begin{equation}
{ F(t) = (F_0-F_\mathrm{disk}) \exp\left[-{A} {\exp\left(-\frac{t-T_0}{b}\right)}\right]+F_\mathrm{disk}},
\end{equation}
where $T_0$ is the initial time at which the flux growth starts, $A$ is a dimensionless
constant, $b$ is a constant coefficient corresponding to the characteristic time of
the flux change, $F_0, F_\mathrm{disk}$ are constant fluxes corresponding to  contributions
from the neutron star and the disk. The choice of such a function instead of, say, a polynomial
approximation is also convenient because it provides the desirable
asymptotics $F(t\ll T_0)\to F_\mathrm{disk}$, $F(t\gg T_0)\to F_0$.

The plots for the best-fit exponential flux growth in the 4--12 and 12--25 keV ranges
after the main Her X-1 turn-on are shown in Figure \ref{fig10}
, respectively (see parameters in Table \ref{tab:tbl3}). It is seen that the model curve
in general well describes the observed sharp flux increase during the opening of the neutron
star by the precessing disk.

\begin{table}[t]
  \caption{Parameter values of the exponential growth model (4)}
  \centering%
  \resizebox{\columnwidth}{!}{
    \renewcommand{\arraystretch}{1.5}
    \begin{tabular}{|c|c|c|c|c|}\hline
      Energy range, keV & $A$  & $b$, d  & $F_\mathrm{disk}$, counts/s \\\hline
      4--12  & $3.41\pm0.07$ & $0.048\pm0.001$ & $2.52\pm0.05$ \\\hline
      12--25 & $1.82\pm0.18$ & $0.049\pm0.002$ & $0.54\pm0.17$ \\\hline
    \end{tabular}
  }
  \label{tab:tbl3}
\end{table}

The obtained parameters of the phenomenological model (4) are insufficient for unambiguous
conclusions about a detailed physical structure of the scattering corona of the disk. 
However, the model allows us to trace the general behavior of the  light curve in both energy
ranges, which is well described  by scattering of radiation in the inhomogeneous corona of
the accretion disk.
The parameter $b$, responsible for the flux increase rate, has close values in both energy
ranges, while all the difference in the flux growth is encoded in the time-independent
parameter $A$ (see Figure \ref{fig12}).

The X-ray hardness plot in the bottom panel of Figure \ref{fig12} shows that during
the phase of exponential flux growth the light curve behaves differently in different
energy ranges, but the saturation of the flux increase occurs simultaneously in both ranges.
This can also be seen in Table \ref{tab:tbl2}, where the value of parameter $b$ is similar
in different energy ranges, while the parameter $A$ is very different. The harder emission
up to the main turn-on is due to the preferential scattering of hard photons in a hot
corona above  the X-ray heated atmosphere of the optical star, as expected in the model
of Her X-1 \citep{2021A&A...648A..39S} .

In Figs. \ref{fig3} and \ref{fig10}
in the post-eclipse light curve, a short peak before the onset of quasi-exponential growth
near the orbital phase 0.18 (interval \#33 in Figure \ref{fig6}) attracts attention.
The peak is more pronounced in the hard range, which can also be seen in the spectrum.
It is seen in Figure \ref{fig9} that in the 12--25 keV energy range, the pulsations from
the neutron star appear slightly earlier than in the 4--12 keV range, just in interval \#33.
Such a flux behavior can be explained by the fact that the registered radiation at this
orbital phase may arrive not only scattered or re-reflected on the accretion disk, but
also on the hot corona above the heated atmosphere of the optical star. It is known
that the X-ray reflection albedo 
increases with the photon energy \citep{2011AstL...37..311M}, which leads to a more pronounced
reflected 12--25 keV pulse emission in  interval \#33 in Figure \ref{fig6}. However, we do not
consider this moment to be the beginning of the main turn-on, since the neutron star must be
still obscured by the tilted precessing accretion disk. Of course, to confirm that
the appearance of this hard peak is spurious, new observations of Her X-1 at this phase
are needed.

\section{Conclusion}
\label{sect:concl}

We present results of the Mikhail Pavlinsky ART-XC telescope observations of the eclipsing
Her X-1/HZ Her X-ray binary system in the end of the second low state near the moment of
the main source turn-on in the interval MJD = 60180.89 -- 60182.87. {A sine-like modulation
of non-pulsating X-ray emission in the off state with a characteristic time of 0.87 days
may be explained by tidal wobbling of the precessing accretion disk around the neutron
star suggested by \citep{2006AstL...32..804K}}. 

The 4--12 keV light curve shows a brief appearance of X-ray pulsations in the off state
near the orbital phase 0.8 at the maximum of the proposed nutation variability, which may
temporarily open the central source before
the main-on \citep[see Figure \ref{fig6} in][]{2006AstL...32..804K}.
In our observations, the Her X-1 main-on is found to occur near the orbital phase 0.2,
which is close to the phases 0.25 and 0.75 where the tidal nutation of the precessing
accretion disk is mostly pronounced \citep{1982ApJ...262..294L, 1982ApJ...260..780K}.
However, main turn-on times in Her X-1 have not always be found near these
phases \citep{2010ApJ...715..897L,2020ApJ...902..146L}, and the model of tidal wobbling
of accretion disk should be further checked by future observations.

A residual X-ray emission during the X-ray eclipse has been measured. The observed
flux is consistent with a model of scattered radiation in an optically thin hot corona
above the X-ray-illuminated photosphere of the optical star that was detected in
Her X-1 observations with the SRG/eROSITA telescope
\citep{2021A&A...648A..39S}.

We analyzed the X-ray light curve after the main turn-on of Her X-1 and showed that
it is possible to approximate the flux growth taking into account only scattering and
time evolution of the column density associated with change in the number of particles
along the line of sight when the source is being opened by the outer edge of the
precessing accretion disk.


\begin{acknowledgements}

This work is based on observations with the Mikhail Pavlinsky ART-XC telescope,
hard X-ray instrument on board the SRG observatory. The SRG observatory was created
by Roskosmos in the interests of the Russian Academy of Sciences represented by its
Space Research Institute (IKI)
in the framework of the Russian Federal Space Program, with the participation of
Germany. The ART-XC team thanks the Roscosmos State Corporation,
the Russian Academy of Sciences, and Rosatom State Corporation for supporting
the ART-XC telescope, as well as the JSC Lavochkin Association and partners
for manufacturing and running the Navigator spacecraft and platform.
The work was supported by the Minobrnauki RF grant 075-15-2024-647.
\end{acknowledgements}

%
%

\bibliographystyle{aa}
\bibliography{aa55103-25.bib}

\end{document}